\documentclass[aps,twocolumn,showpacs,preprintnumbers,amsmath,amssymb,showkeys,prb]{revtex4-1}

\usepackage{graphicx}
\usepackage{color}
\usepackage{amsmath,amssymb,mathrsfs}
\def\vk{{\bf k}}
\def\bra{\langle}
\def\ket{\rangle}
\newcommand{\bk}{\mathbf{k}}
\def\bR{{\bf R}}
\newcommand{\tchikm}{|\tilde{\chi}^{\bR\sigma}_{\bk m}\rangle}
\newcommand{\wfkm}{|W^{\bR\sigma}_{\bk m}\rangle}
\def\vr{{\bf r}}
\def\vq{{\bf q}}
\def\vG{{\bf G}}

\begin{document}

\title{Screened Coulomb interaction calculations: cRPA implementation 
and applications to dynamical screening and self-consistency in uranium dioxide and cerium}

\author{Bernard Amadon}
\email{bernard.amadon@cea.fr}
\affiliation{CEA, DAM, DIF, F-91297 Arpajon, France}
\author{Thomas Applencourt}
\altaffiliation{Present address:  Univ Toulouse, CNRS IRSAMC, Lab Chim \& Phys Quant, Toulouse, France}
\affiliation{CEA, DAM, DIF, F-91297 Arpajon, France}

\author{Fabien Bruneval}
\affiliation{CEA, DEN, Service de Recherches de M\'etallurgie Physique, F-91191 Gif-sur-Yvette, France}


\begin{abstract} 
We report an implementation of the constrained Random Phase Approximation (cRPA) method within the Projector Augmented-Wave framework. It allows for the calculation
of the screened interaction in the same Wannier orbitals as our recent DFT+$U$ and DFT+DMFT implementations.
We present calculations of the dynamical Coulomb screened interaction
in uranium dioxide and $\alpha$ and $\gamma$ cerium on Wannier functions.
We show that a self-consistent calculation of the static screened interaction in DFT+$U$ together with a consistent Wannier basis is mandatory
for $\gamma$ cerium and uranium dioxide.
We emphasize that a static approximation for the screened interaction in $\alpha$ cerium is too drastic.
\end{abstract}

\maketitle

\section{Introduction}

Because of the limited accuracy of available functionals, Density Functional Theory (DFT) fails for a large number of correlated systems.
There are numerous examples for which DFT cannot describe neither their ground state properties, nor their excitation properties.
Thus, in order to describe many-body effects arising in the strongly correlated systems containing for instance transition elements or $f$ electrons,
theories were designed to take into account the interaction among correlated orbitals explicitly. The DFT+$U$ method\cite{Anisimov1991} or
the combination of DFT with Dynamical Mean Field Theory method (DFT+DMFT) \cite{Lichtenstein1998,Kotliar2006} have been successfully applied
to a large number of systems in the last twenty years. 
In particular, these methods have been particularly useful to study the ground
state and the photoemission spectra of Mott insulators such as bulk actinide\cite{Dudarev1997,Dudarev1998,Laskowski2004,Sun2008,Jomard2008,Dorado2009,Yin2011,Amadon2012} and lanthanide oxide\cite{Andersson2007,Dasilva2007,Loschen2007,Pourovskii2007a,Amadon2012,Jiang2012}. 
For instance, to date DFT+DMFT is the only method
to give a good description of photoemission spectra of both $\alpha$ and $\gamma$ cerium 
\cite{Held2001,McMahan2003,Haule2005,Amadon2006,Streltsov2012,Bieder2013a}.
However, in these frameworks and applications, the interaction among correlated orbitals, named $U$, remained most often an input parameter.

As a consequence, there is a stringent need to calculate the magnitude of the interaction $U$, in order to recover a truly \textit{ab initio} scheme.
Methods were then proposed to evaluate $U$ from first-principles. The constrained Local Density Approximation (cLDA) method\cite{Gunnarsson1989,Anisimov1991} deduces the 
value of $U$ from the variation of energy with respect to the number of correlated electrons on an atom. Later,
Cococcioni {\it et al} \cite{Cococcioni2005}  generalized this method to a non basis dependent scheme.
Finally, the constrained RPA method\cite{Aryasetiawan2004,Aryasetiawan2006} uses the linear response theory to compute 
the value of the screened interaction.  Screening processes corresponding to electron hole transitions among the correlated orbitals are 
however excluded from the calculation.
Indeed an exact many-body scheme would already contain all the screening processes associated to the degrees of freedom involved in the calculation.
Therefore, some transitions have to be disregarded to avoid double-counting.
The cRPA scheme very clearly defines which screening processes have to be taken into account. 
The method has been implemented in several electronic structure codes using the LMTO\cite{Aryasetiawan2004,Aryasetiawan2006}, 
FLAPW\cite{Kutepov2010,Sasioglu2011,Vaugier2012}, FPLMTO\cite{Miyake2008a},Plane Wave\cite{Shih2012,Nomura2012} and Projector Augmented-Wave (PAW)\cite{Nomura2012} methods and 
applied to different systems in the last few years\cite{Miyake2008a,Miyake2009,Kutepov2010,Sasioglu2011,Nomura2012,Vaugier2012,Shih2012,Shih2012a,Sakuma2013}.
An important point emphasized in several works  is that 
the calculation of the cRPA screened interaction based on a previous DFT calculation depends crucially on the definition of a many-body model\cite{Vaugier2012,Sakuma2013}.

The many-body model is defined by a set of local orbitals together with the interactions among them.
A cRPA calculation of a model would require first the definition of a set of local orbitals, and second a consistently calculated screened interaction.
Whereas the choice of the angular momenta of the selected orbitals unambiguously defines the angular part of the local orbitals,
the definition of the radial part is more subjected to variation.
Generally, it relies on the use of localized Wannier functions, which are built as a unitary transform of Kohn-Sham orbitals in an energy window\cite{Wannier1937,Marzari1997}.
The key point is to construct the model specific cRPA screened interaction.

It is especially important to study this dependence as a function of the localization of the Wannier orbitals. Indeed, correlated orbitals used in DFT+$U$ and DFT+DMFT can be formulated  as Wannier orbitals with different energy windows\cite{Lechermann2006,Amadon2008,Amadon2012} depending on implementation choices. 
Thus the coherence of the DFT+$U$/DFT+DMFT calculations and of the cRPA calculation can only be guaranteed if both methods
use the same Wannier functions. 
Implementations of DFT+$U$ in popular codes very often use atomic orbitals\cite{Shick1999,Bengone2000,Amadon2008a}, whereas implementations of DFT+DMFT use Wannier functions\cite{Anisimov2005,Lechermann2006,Amadon2008,Haule2010,Amadon2012,Granas2012}.
It is thus expected that the value of $U$ used in DFT+$U$ and DFT+DMFT should differ.
Some works indeed discuss the calculation of $U$ for a given energy window\cite{Aryasetiawan2006,Miyake2008a,Miyake2008b,Miyake2009,Vaugier2012} but
there are no calculation of a screened interaction in the same basis as the one used in DFT+$U$ codes.
Even for DFT+DMFT calculation, the definition of an energy window is especially important for systems 
with entangled correlated bands\cite{Aryasetiawan2006,Miyake2009}.

Moreover and especially for very localized systems --- such as Mott insulators ---, self-consistent calculations over $U$\cite{Karlsson2010a,Shih2012a} are desirable because the erroneous LDA or Generalized Gradient Approximation (GGA) band structures cannot
correctly describe the screening in these systems.
The coherence of basis between the DFT+$U$ and the cRPA calculation
is then of the utmost importance.
This is in particular the case for cerium and uranium dioxide for which 
no self-consistent calculations of $U$ exist.

In this work, we report an implementation of the cRPA method using the PAW method in 
the \textsc{Abinit} package\cite{Torrent2008,Giantomassi2011,Gonze2009}. 
The implementation is versatile enough to allow for the calculation
of screened interaction in both the same Wannier basis as our recent DFT+DMFT implementation\cite{Amadon2012}
and in atomic orbitals and as in our DFT+$U$ implementation\cite{Amadon2008a}.
Then, we show that the self-consistent DFT+$U$ calculation of the screened interaction in the 
cRPA method in strongly correlated systems is essential to describe the static and dynamical screened interaction.
We exemplify our study with two important applications: UO$_2$ and cerium.

\section{The constrained RPA method}

As the cRPA method is described in details elsewhere \cite{Aryasetiawan2004,Aryasetiawan2006,Vaugier2012,Sakuma2013}, 
we only sketch the most important points here.

The screened interaction is in general a four index matrix that is defined as
\begin{equation}
  U^{\sigma,\sigma'}_{m_1,m_3,m_2,m_4}(\omega)=\bra m^\sigma_1 m^{\sigma'}_3 |\varepsilon_r^{-1}(\omega) v | m^\sigma_2 m^{\sigma'}_4 \ket
\label{eq:ummmm}
\end{equation}
with $v$ the bare Coulomb interaction and $m_1, m_2, m_3, m_4$ the indices of the correlated orbitals.

In this work, the use Projected Wannier functions as defined in Refs. \onlinecite{Amadon2008} and \onlinecite{Amadon2012}. Similar Wannier functions have been used in cRPA calculations\cite{Vaugier2012}.
We first introduce the auxiliary wavefunctions $\tchikm$ as
\begin{equation}
\tchikm\equiv\sum_{\nu\in {\cal W}}
|\Psi^\sigma_{\bk\nu}\rangle
\langle\Psi^\sigma_{\bk\nu}|\chi^{\bR}_{\bk m}\rangle.
\label{eq:tchi}
\end{equation}
For a given atomic site $\bR$, we call
$|\chi^{\bR}_{\bk m}\ket$ the Bloch transform of isolated atom Kohn Sham orbitals with projected angular momentum $m$. 
$|\Psi^\sigma_{\bk\nu}\ket$ are Kohn-Sham orbitals for k-point $\bk$, band index $\nu$ and spin $\sigma$.
$\tchikm$ is thus a weighted sum of Kohn-Sham orbitals. 
This sum extends over a given number of Kohn Sham orbitals that can be defined by an index range or alternatively by an energy window  ${\cal W}$.
The orthonormalization of $\tchikm$ leads to well defined Wannier functions $\wfkm$, unitarily related
to $|\Psi^\sigma_{\bk\nu}\ket$.
In the limit of a large number of Kohn Sham bands, the projection in Eq. (\ref{eq:tchi}) becomes complete and the Wannier functions  $\wfkm$  become equivalent
to atomic orbitals $|\chi^{\bR}_{\bk m}\ket$.

The definition of $U$ is very much similar to the Coulomb integrals used in quantum chemistry, but with the screening of the frequency dependent cRPA dielectric matrix, $\varepsilon_r(\omega)$.
This dielectric matrix can be expressed as a function of the cRPA non-interacting polarizability $\chi_0^r(\omega)$ and the bare interaction $v$ as (in the matrix notation)
\begin{equation}
  \varepsilon_r(\omega) = 1 - v \chi^r_0(\omega).
\end{equation}

$\chi^r_0$
contains all electron-hole screening processes
except the ones that are internal to the correlated orbitals of the model.
It can be conveniently written as
\begin{multline}
\chi_0^{{r}}(\vr,\vr',\omega) =
 \sum_{\vk,\vk',n,n',\sigma}
    \psi^{\sigma *}_{n \mathbf{k}}(\vr) \psi^\sigma_{n' \mathbf{k'}}(\vr)
    \psi^{\sigma *}_{n' \mathbf{k'}}(\vr') \psi^\sigma_{n \mathbf{k}}(\vr') \\
\times 
   w(\vk,\vk',n,n',\sigma) \frac{f^\sigma_{n'\mathbf{k'}}-f^\sigma_{n \mathbf{k}}}{\epsilon^\sigma_{n'\mathbf{k'}}-
     \epsilon^\sigma_{n\mathbf{k}}+\omega+i\delta}.
\label{eq:pola}
\end{multline}
In Eq.~(\ref{eq:pola}), $n,n'$ are band indices, $\vk,\vk'$ are k-points in the Brillouin Zone and $f_{n\vk}$ is the occupation number
for band $n$, spin $\sigma$ and k-point $\vk$.

If the correlated bands in the model are completely isolated from the other ones, then we can assume that\cite{Aryasetiawan2004,Aryasetiawan2006}
\begin{equation} 
   w(\vk,\vk',n,n',\sigma)=0
\label{eq:pola0}
\end{equation} 
when ($n\mathbf{k}$)  and ($n'\mathbf{k}'$) are both correlated bands and $w=1$ otherwise.
For example, a model could define correlated orbitals as Wannier orbitals constructed only from the $f$ bands.
Nevertheless, with this specific choice,
the Wannier orbitals have some weight on other orbitals: Oxygen-$p$ for oxides\cite{Lechermann2006} or $spd$ for pure metals.
The intensity of this weight
depends on the hybridization of $f$ orbitals with the other orbitals. This last definition of correlated orbitals is 
not the one used in most implementations of DFT+$U$ in modern codes\cite{Shick1999,Bengone2000,Amadon2008a}.
In these implementations, correlated orbitals are most often atomic orbitals which thus corresponds to
Wannier functions $\wfkm$ for a large window of energy\cite{Amadon2012}.

If the bands are completely entangled or if one defines Wannier functions $\wfkm$ from a larger energy window, then the preceding assumption of Eq. (\ref{eq:pola0}) cannot be made\cite{Aryasetiawan2006,Miyake2009} and some authors have proposed the more general assumption\cite{Shih2012,Sakuma2013}:
\begin{multline}
  w(\vk,\vk',n,n',\sigma)= 1- \left[\sum_{m_1} |\bra\Psi^\sigma_{n\mathbf{k}}  | W^\sigma_{m_1\vk} \ket|^2\right] \\
                   \times
                    \left[\sum_{m_2} |\bra\Psi^\sigma_{n'\mathbf{k'}}| W^\sigma_{m_2\vk'}\ket|^2\right] .
\label{eq:weightpola}
\end{multline}
If the correlated bands are not entangled and if the Wannier functions are defined from these correlated bands only, then
Eq.~(\ref{eq:weightpola}) simply reduces to Eq.~(\ref{eq:pola0}). Fully screened coulomb interaction $W$ corresponds to $w=1$.

In this work, we use an implementation of the calculation of the dielectric function in PAW\cite{Arnaud2000,Shishkin2007,Giantomassi2009,Giantomassi2011}.
From the screened interaction expressed in the Kohn Sham basis, we compute the screened interaction in Eq.~(\ref{eq:ummmm}),
using the Wannier functions as defined in Refs.~\onlinecite{Amadon2008} and~\onlinecite{Amadon2012}.
The weight of Wannier functions necessary for Eq.~(\ref{eq:weightpola}) is evaluated within PAW following Ref. \onlinecite{Amadon2008}.

Then, the values of the famous Hubbard $U$ and Hund $J$ are simply extracted
by taking the average among the considered localized orbitals:
\begin{equation}
U=\frac{1}{4} \sum_{\sigma,\sigma'} \frac{1}{(2l+1)^2} \sum_{m_1=1}^{2l+1}\sum_{m_2=1}^{2l+1} U^{\sigma,\sigma'}_{m_1,m_2,m_1,m_2}
\label{eq:udef}
\end{equation}
\begin{equation}
J=\frac{1}{4} \sum_{\sigma,\sigma'} \frac{1}{(2l+1)(2l)} \sum_{m_1=1}^{2l+1}\sum_{m_2=1 (m_2\neq m_1)}^{2l+1} U^{\sigma,\sigma'}_{m_1,m_2,m_2,m_1}
\label{eq:jdef}
\end{equation}
Note that this definition of $U$ is also sometimes referred to as the $F^0$ Slater integral.
We emphasize that this definition is different from the average of the diagonal elements of
the Coulomb interaction matrix $U_{\rm diag}=\frac{1}{2}\sum_{\sigma}\frac{1}{2l+1} \sum_{m_1=1}^{2l+1} U^{\sigma,\sigma}_{m_1,m_1,m_1,m_1}$ (see e.g Ref. \onlinecite{Shih2012}).
In particular diagonal elements are usually larger, and thus $U_{\rm diag}>U$.
However our definition is coherent with the $U$ used in the DFT+$U$  approach\cite{Liechtenstein1995} and physically describes
the average interaction between electrons in all orbitals.

Appendix \ref{sec:Udetails} gives the details of the implementation and the peculiarities of the PAW formalism for the calculation of
$U$.
Appendix \ref{sec:appendix} gives a benchmark of our implementation with respect to recent calculations
on SrVO$_3$.

\section{Computational details}
The calculation are performed within the Projector Augmented-Wave (PAW) method as implemented in \textsc{Abinit}\cite{Gonze2009,Torrent2008,Giantomassi2011}. 
The valence states include
$2s,2p$ for oxygen, $5s,5p,4f,6s,5d$ for cerium and $6s,6p,5f,7s,6d$ for uranium respectively.
Two projectors per angular momentum are used, and completeness of the projector basis is checked by increasing their number.
The parameters of calculation are chosen such that the precision on the static values of $U$ and $J$ is better than 0.2 eV. For UO$_2$,
we thus use a 4x4x4 k-point grid, and energy cutoffs for the wavefunctions, the dielectric function and the bare Coulomb
interaction are respectively 15 Ha, 5 Ha, and 35 Ha. 100 bands are sufficient for the calculation of the polarisability.
For cerium, we use a 8x8x8 k-point grid and energy cutoffs for the  wavefunction, the dielectric function and the bare Coulomb
interaction are respectively 15 Ha, 10 Ha and 35 Ha (for large values of the volume, a 4x4x4 k-point grid was sufficient). For the static screened exchange $J$ in cerium, a value of 120 Ha was however necessary but a 4x4x4 k-point grid is sufficient as well as 100 bands for the calculation of the polarisability. This high value of the cutoff originates from the calculation
of oscillator matrix elements in the PAW formalism (see appendix \ref{sec:Udetails}).
A smearing of the Kohn Sham occupations of 0.1 eV is used.
For all systems, experimental structural parameters  are used: 5.47 \AA \; for UO$_2$, 4.83 \AA \; for $\alpha$ cerium, 
and 5.16 \AA\; for $\gamma$ cerium.

Unless specified, all DFT+$U$ calculations use the Full Localized Limit (FLL) double
counting correction\cite{Liechtenstein1995}.
A discussion on the role of the double counting correction is given
in appendix \ref{app:dc}.
For UO$_2$, DFT+$U$ are performed for simplicity in the ferromagnetic configuration, 
which requires a symmetry breaking\cite{Larson2007} and we use the
correlated $f$ density matrix found in Ref. \onlinecite{Dorado2009}.

\section{Definition of the models for uranium dioxide and cerium}

The goal of this section is to define models of correlation for uranium dioxide and cerium.
For each model, one thus defines an energy window --- that encompasses at least the bands which have
the same main character as the selected orbitals.
From the definition of the window energy, Wannier functions of the correlated orbitals
are built according to the scheme of Ref.~\onlinecite{Amadon2008}.
From the choice of the correlated orbitals, the cRPA polarizability is built by excluding some
screening channels corresponding to correlated orbitals.
We distinguish different ways to exclude the screening according to Eq.~(\ref{eq:pola0}) or to Eq.~(\ref{eq:weightpola}).

\subsection{UO$_2$}

\begin{figure}[t]
\centering
 {\resizebox{8cm}{!}{\includegraphics{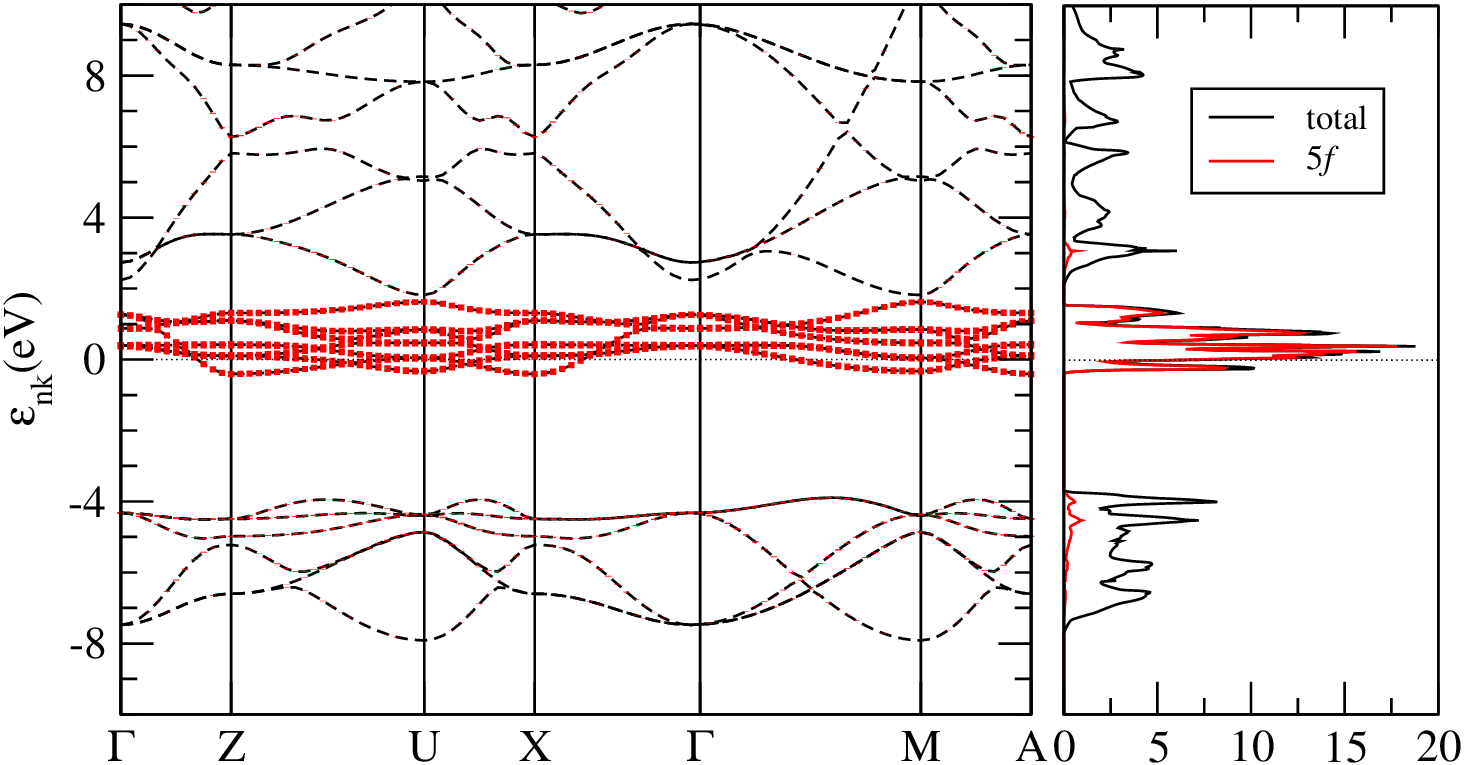}}}
  \caption{UO$_2$ band structure for the DFT-LDA non magnetic solution. 
  The Fermi level is set to zero.
  The width of the bands (in red) is proportional to the amplitude of the $f$ character for each band.}
  \label{fig:uo2.spectra}
\end{figure}

\begin{table}[b]
\centering
\begin{tabular}{cccc}
\hline \hline 
  Models           &     Excluded            & \multicolumn{2}{c}{Wannier functions}                \\
                   &    screening channels   & defined                & corresponding               \\
                   &                         & with bands             & energy window (eV)          \\
 $f$               &  $f$                    &  $f$          &    [-1, 1.7 ]          \\
 $fp$              &  $f$, O-$p$             &  $f$, O-$p$   &    [-8, 1.7 ]          \\
 $f-fp$ (a)        &  $f$                    &  $f$, O-$p$   &    [-8, 1.7 ]          \\
 $f-fp$ (b)        &  $f$-Wannier weight     &  $f$, O-$p$   &    [-8, 1.7 ]          \\
 $f$-ext (b$_a$)   &  $f$-Wannier weight     &    5-28       &    [-8, 17.0]          \\
 $f$-ext (b$_b$)   &  $f$-Wannier weight     &    5-38       &    [-8, 30.0]          \\
 $f$-ext (b$_c$)   &  $f$-Wannier weight     &    5-48       &    [-8, 40.0]          \\
\hline \hline
\end{tabular}
\caption{Different models for the description of correlation in UO$_2$. From the top to the bottom, the $f$ Wannier function
are expected to be more and more localized. For the first three model (resp. last four models), 
Eq. (\ref{eq:pola0}) (resp Eq.~(\ref{eq:weightpola})) is used to compute $\chi_0^{r}$.
The three models $f$-ext (b$_a$) (b$_b$) (b$_c$) use a fixed number of bands (respectively 28, 38 and 48) that corresponds to the energy windows given.}
\label{tab:modelsuo2}
\end{table}

Figure~\ref{fig:uo2.spectra} represents the
LDA band structure of UO$_2$.
The O$p$-like bands are located below the Fermi level in the energy window [-8 eV, -4 eV].
Near the Fermi level, bands have mainly a U $f$ character and are non entangled in this LDA non magnetic calculation.
As a consequence, one can define several models following the literature\cite{Sakuma2013}, as listed below.
We give their energy window and screening channel excluded from the polarizability in Tab.~\ref{tab:modelsuo2}:

\begin{itemize}
\item[]$f$ model: The model is built from the U$f$-like bands only.
\item[]$fp$ model: The model is built from the U$f$-like and O$p$-like bands.
\item[]$f-fp$ model (a): As in the $fp$ model,  Wannier functions are built from the U$f$-like and O$p$-like bands. However, only
the $f$ bands transitions are removed from the polarizability, using Eq.~(\ref{eq:pola0}). It is equivalent to say that
the constrained polarizability is built from Eq.~(\ref{eq:weightpola}) with Wannier orbitals constructed from U$f$-like bands only. Thus from
an {\it ab initio} point of view, this scheme is not coherent.
\item[]$f-fp$ model (b):  Wannier functions are also built from the U$f$-like and O$p$-like bands. Nevertheless, in this case, the cRPA
polarizability is computed using Eq. \ref{eq:weightpola} in Eq. \ref{eq:pola} \cite{Sasioglu2012,Shih2012}.
This is a more general way of doing because it is applicable to any system, even when bands are entangled.
Furthermore the Wannier functions and the cRPA polarizability are here consistently defined.
\item[]$f$-ext model (b):  The same as $f-fp$ model (b)  but Wannier functions are defined with more extended 
window of energy that are precised in Tab.~\ref{tab:modelsuo2}.
\end{itemize}

\subsection{Cerium}

\begin{figure}[t]
\centering
 {\resizebox{8cm}{!}{\includegraphics{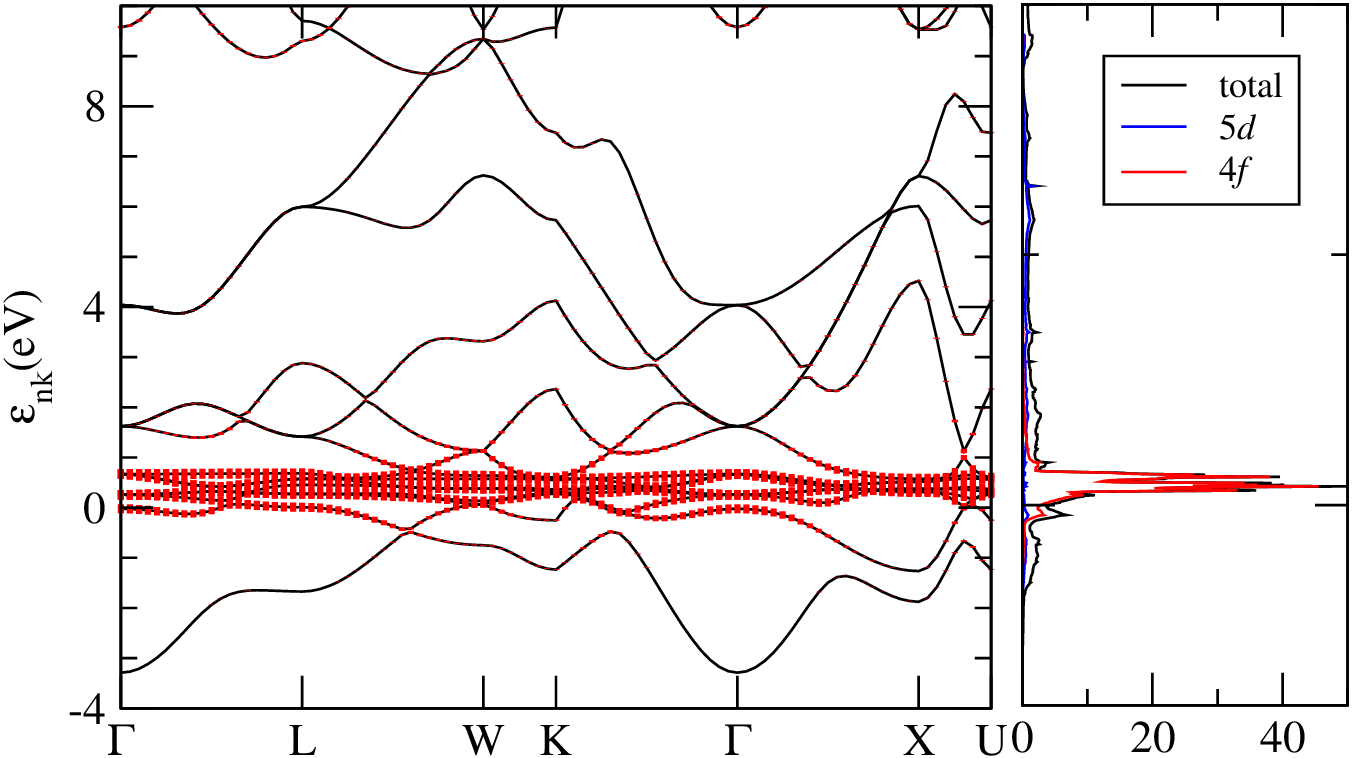}}}
  \caption{Band structure and density of states of $\gamma$ cerium in DFT-LDA.
}
  \label{fig:ce.spectra}
\end{figure}

The LDA band structure of $\gamma$ cerium is given in Fig.~\ref{fig:ce.spectra}.
One can see that the $f$ bands are largely entangled with $s$, $p$, and $d$ bands,
The $f$, $fp$ and $f-fp$ (a) models we defined for UO$_2$ cannot be applied here.
We give in Tab.~\ref{tab:modelce} the list of models that we will use in the next section.

The four models $f$-ext (b$_1$) (b$_2$) (b$_3$) (b$_4$) use a fixed number of bands (respectively 20, 30, 40 and 50) that corresponds to the energy windows given.
The $f$-(${\cal W}_1$) models are built to select an energy window ${\cal W}_1$ to remove the $f$ bands contribution approximatively in the polarizability.
The $fd_{t_{\rm 2g}}$-ext model uses Wannier $f$ functions  constructed from the specified energy window and the excluded bands for the polarisability are the $f$ and $d_{t_{\rm 2g}}$ bands.
In this last model, as all bands are entangled, we choose to remove the 7 $f$ bands and the 3 bands that are located just above as they are mainly of $d_{t_{\rm 2g}}$ character
and are lower in energy than the $d_{e_{\rm g}}$ orbitals.

\begin{table}[b]
\centering
\begin{tabular}{cccc}
\hline \hline 
  Models               &     Excluded              & \multicolumn{2}{c}{Wannier functions}                \\
                       &    screening channels     & defined                & corresponding               \\
                       &                           & with bands             & energy window (eV)          \\
 $f$-ext (b$_1$)       &  $f$-Wannier weight       & 1-20       &  \hspace{0.5cm}   [-24, 27]      \\
 $f$-ext (b$_2$)       &  $f$-Wannier weight       & 1-30       &  \hspace{0.5cm}   [-24, 47]      \\
 $f$-ext (b$_3$)       &  $f$-Wannier weight       & 1-40       &  \hspace{0.5cm}   [-24, 57]      \\
 $f$-ext (b$_4$)       &  $f$-Wannier weight       & 1-50       &  \hspace{0.5cm}   [-24, 67]      \\
 $f$-(${\cal W}_1$) $\alpha$  &  [-0.8,0.4]               & 1-20       &  \hspace{0.5cm}   [-24, 27]      \\
 $f$-(${\cal W}_1$) $\gamma$  &  [-0.63,0.37]             & 1-20       &  \hspace{0.5cm}   [-24, 27]      \\
 $fd_{t_{\rm 2g}}$-ext &  $fd_{t_{\rm 2g}}$ bands  & 1-20       &  \hspace{0.5cm}   [-24, 27]      \\
\hline \hline
\end{tabular}
\caption{ 
Different models for the description of correlation in cerium (see text)
}
\label{tab:modelce}
\end{table}

\section{UO$_2$: Results and discussion}
In this section, we present the static and dynamical cRPA screened interaction in uranium dioxide.
The third subsection is devoted to the self-consistent calculation  of the static screened interaction.

\subsection{Static screening}

Tab. \ref{tab:uo2.static} gives the static values of bare $v$, fully screened $W$, and cRPA value $U$
of the direct and exchange interactions as defined in Eq. (\ref{eq:udef}) and (\ref{eq:jdef}).

\subsubsection{Limiting cases: the bare and the fully screened interactions}

We first focus on the bare value of the interaction as a function of the definition of the Wannier function.
Here the screening is completely neglected, i.e. the dielectric matrix is set to 1 in Eq.~(\ref{eq:ummmm}).
As expected, the larger the window of energy used to define Wannier functions (from the $f$ to the $f$-ext model), the larger the value
of the bare interactions $U$ and $J$, ranging from 16.0 eV to 18.1 eV.
Indeed, the larger the energy window, the more localized the Wannier functions.

For the fully screened interaction, the value of the interaction is much reduced by the screening.
However, the same variation is logically observed as a function of the energy window
used to define Wannier functions.

\subsubsection{cRPA interaction}

We compare now the value of the cRPA interaction for the different models.
We take as a reference the value within the $f$ model, namely $U^{(f)}$= 3.4 eV.   
For the $f-fp$ (a) model, the polarizability is computed in the same way, but the Wannier functions
are more localized. Consequently the value for this model is slightly larger
$U^{f-fp (a)}$=3.7 eV.
The relative increase of the interaction when one goes
from the $f$ model to the $f-fp$ (a) is the same for the bare, fully screened and cRPA interactions.

Let us now compare $f-fp$ (a) and $f-fp$ (b) models to highlight the impact of the 
change in the definition of the polarizability, for a fixed Wannier function.
One remarks that the value for $f-fp$ (b) is considerably smaller than for the $f-fp$ (a) model. 
The decrease originates 
from the large hybridization between oxygen and uranium which creates a residual oxygen contribution near
the Fermi level and thus a very efficient metallic screening. A similar effect has been observed in transition metal oxides\cite{Sakuma2013}.

The model $f$-ext (b) corresponds to an even more localized $f$-Wannier function. But the main effect is that the weight of the Wannier function
on Kohn Sham bands around the Fermi level decreases. Thus the remaining screening channels at the Fermi level are more important, it creates 
a larger metallic screening and the value of $U$ is thus even more reduced.
This model is the most relevant because it can be applied for entangled bands, and is fully coherent with modern DFT+DMFT implementations.

Finally, the $fp$ model is based on the same Wannier functions as the $f-fp$ (ab) models but the screening is much more
reduced because the transitions internal to $f$-like and $p$-like bands are removed.

\begin{table}[b]
\centering
\begin{tabular}{ccccc}
\hline \hline 
	      &   model                  &  $U$ (eV) & $J$ (eV)\\
\hline
$v$             & $f$                    &  16.0  & 0.5  \\
$v$             & $fp$ or $f-fp$(a,b)    &  17.1  & 0.5  \\
$v$             & $f$-ext (b$_a$)            &  18.1  & 0.5  \\
\hline
$W$             & $f$                    &   0.20 & 0.3  \\
$W$             & $fp$ or $f-fp$(a,b)    &   0.21 & 0.4  \\
$W$             & $f$-ext (b$_a$)            &   0.23 & 0.4  \\
\hline
$U^{\rm cRPA}$             & $f$                    &  3.4  & 0.4  \\
$U^{\rm cRPA}$             & $f-fp$ (a)             &  3.7  & 0.4  \\
$U^{\rm cRPA}$             & $f-fp$ (b)             &  2.0  & 0.4  \\
$U^{\rm cRPA}$             & $f$-ext (b$_a$)        &  1.0  & 0.4  \\
$U^{\rm cRPA}$             & $fp$                   &  6.2  & 0.4  \\
\hline
$U^{\rm cRPA}_{\rm  nsc}$& $f$-ext (b$_a$)       &  5.0  & 0.4  \\
$U^{\rm cRPA}_{\rm  nsc}$& $f$-ext (b$_b$)       &  5.3  & 0.4  \\
$U^{\rm cRPA}_{\rm  nsc}$& $f$-ext (b$_c$)       &  5.5  & 0.4  \\
$U^{\rm cRPA}_{\rm  sc}$& $f$-ext (b$_a$)       &  5.2   & 0.4  \\
$U^{\rm cRPA}_{\rm  sc}$& $f$-ext (b$_b$)       &  5.7   & 0.4  \\
$U^{\rm cRPA}_{\rm  sc}$& $f$-ext (b$_c$)       &  5.7   & 0.4  \\
$U^{\rm cRPA\dagger}_{\rm  sc}$& $f$-ext (b$_a$)   & 5.0 & 0.4  \\
\hline \hline
\end{tabular}
\caption{
Bare ($v$), fully screened ($W$) and cRPA ($U^{\rm cRPA}$) Coulomb interactions for UO$_2$. In cRPA, the screening is computed for the different models described in Tab. \ref{tab:modelsuo2}.
All calculations are done in the non magnetic states for LDA, and ferromagnetic states for LDA+$U$.
The calculation done in the antiferromagnetic configuration
is indicated by $\dagger$. Non self-consistent (nsc) calculations of $U$ use LDA+$U$ with 
$U$=4.5 eV and $J$=0.5 eV.
}
\label{tab:uo2.static}
\end{table}

For the sake of completeness, we have compared LDA versus GGA calculation of $U$. The difference is weak, at most 0.2 eV.
The magnetic state --- non-magnetic or ferromagnetic --- has also a weak effect, below 0.3 eV.
The interest of doing non magnetic calculations is that in this case and as shown on
Fig. \ref{fig:uo2.spectra}, the seven $f$-like bands located near the Fermi level are separated from other bands, so
we can compare rigourously the different models.

As a conclusion of this study, we performed calculations of $U$ in the DFT+$U$ framework with $U$=4.5 eV and $J$=0.5 eV.
DFT+$U$ gives a better description of the band structure of this Mott insulator by opening a gap in agreement with photoemission spectra\cite{Dudarev1998}.
As a consequence, the low energy transitions disappear in the polarizability. Thus,
the screening is less efficient and the value of the cRPA screened interaction is much larger.
It emphasizes the need for a better starting point
than LDA for the cRPA calculation.
In order to fix this issue, we propose a self-consistent procedure\cite{Karlsson2010a}
as discussed in Sec. \ref{sec:Usc}.

\subsection{Dynamical screening}

\begin{figure}[t]
\centering
 {\resizebox{8cm}{!}{\includegraphics{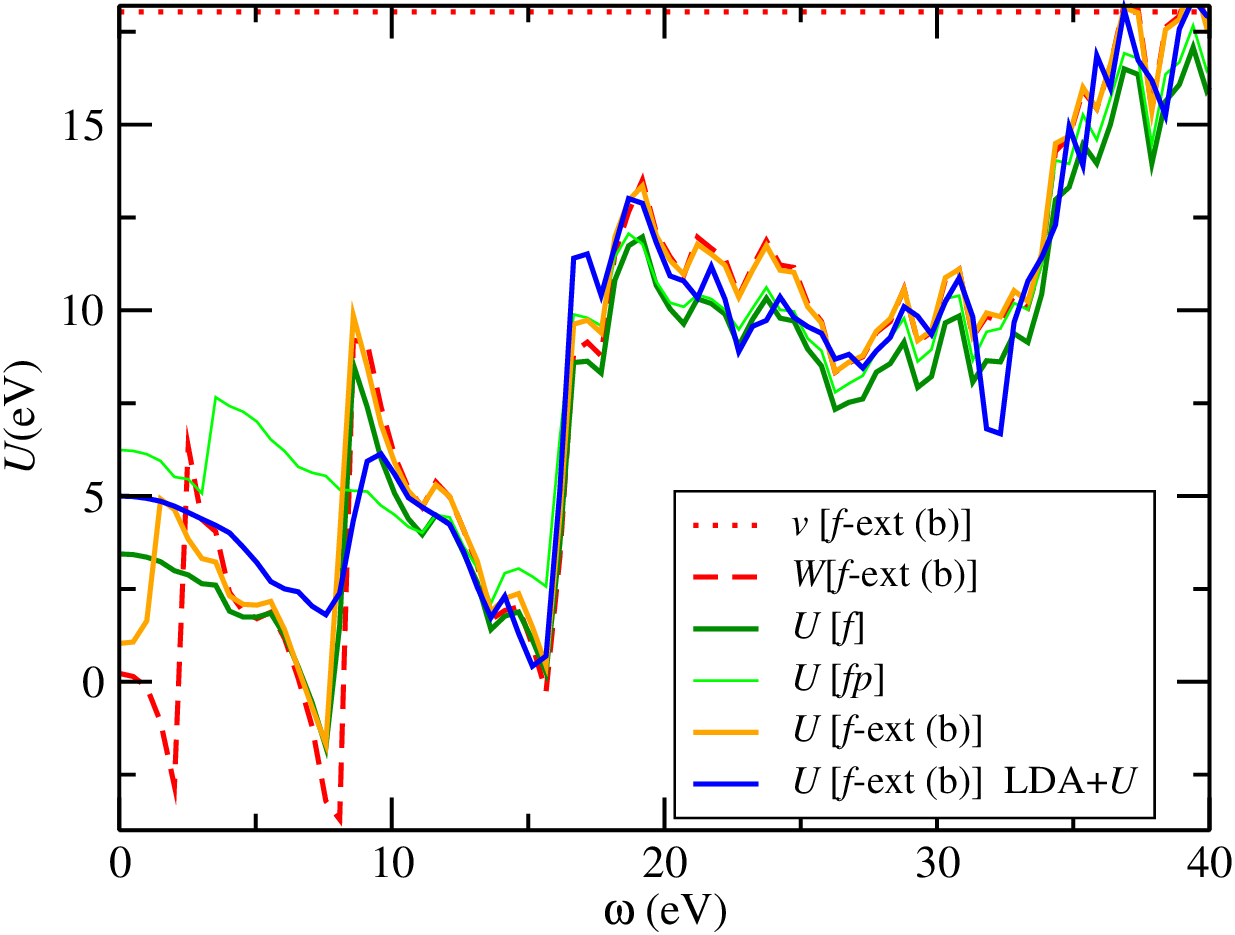}}}
  \caption{Bare, fully screened and cRPA partially screened interactions, for different models, for uranium dioxide. The calculation are performed in the LDA approximation or LDA+$U$ approximation ($U$=4.5 eV and $J$=0.5 eV).}
  \label{fig:uo2.uw}
\end{figure}

We discuss here the frequency dependence of the screened interaction. 
We plot on Fig. \ref{fig:uo2.uw} the screened interactions as a function of the frequency.

\subsubsection{Fully screened interaction.}
We first discuss the frequency dependence of the fully screened interaction. 
For this case, only the interaction computed with the Wannier functions built in the $f$-ext (b) model is shown because the main
features are mainly independent from the Wannier function construction details.
Three peaks are located at 2.5 eV, 8.2 eV and 16.3 eV (called subplasmons in Ref. \onlinecite{Sakuma2013}).
As we will show below by comparing different models, these peaks reflect the shape of the band structure.

\subsubsection{$f$ and $fp$ models}
By comparing $W$ to $U$ computed in the $f$ model, one notices that the peak at 2.5 eV comes from internal
transitions of the $f$ bands. It is coherent with the width of the $f$ bands which --- as seen on Fig. \ref{fig:uo2.spectra} --- is around 2.5 eV.
Similarly, the comparison of the $U$ computed in the $fp$ model, show that the peak at 8.2 eV comes from
the $p-f$ transitions. 

\subsubsection{$f$-ext (b) model}
The cRPA screened interaction as computed in the $f$-ext(b) model exhibits the three peaks also observed in the fully screened interaction W. Similarly to what was observed for transition metal oxides\cite{Sakuma2013}, and as discussed above, it comes from the large hybridization between oxygen and uranium, which produces a residual metallic screening in the cRPA polarizability. As a consequence the first subplasmon is observed, but its amplitude is much lower.
Results for the $f-fp$ (b) model are very similar to the $f$-ext(b) thus have not been reproduced here.

\subsubsection{$f$-ext (b) model in LDA+$U$}
In the DFT+$U$ approximation, a gap is created inside the band structure, thus the first subplasmon is suppressed.
As a consequence, the variation of $U$ as a function of frequency is weaker below 4 eV. 
It is a justification to use a static approximation for the screened interaction.

\subsection{Self-consistent calculation of $U$ }
\label{sec:Usc}

The LDA and LDA+$U$ calculations of the cRPA lead to two different static and dynamical screened interactions. It is thus important
to carry out the calculation of $U$ self-consistently\cite{Karlsson2010a,Shih2012a}:
First, a LDA+$U$ calculation is performed. Then, the band structure and wavefunctions are used in a cRPA calculation 
to obtained a new value of $U$ on a given Wannier function.  Then this value of $U$ is injected into another LDA+$U$ calculation until convergence.

Such scheme can be carried out with our present implementation but in order to guarantee the coherence of the calculation, one needs
to use the same correlated basis for the calculation of the screened Coulomb interaction --- Wannier functions\cite{Amadon2012} --- and the application of the Hubbard correction to the Kohn-Sham hamiltonian --- atomic orbitals\cite{Amadon2008a}. 
As outlined in the Appendix B1 of Ref. \onlinecite{Amadon2012}, it is sufficient in our implementation to use a large number of
bands to define Wannier functions, and a specific choice of the correlated occupation matrix to satisfy the former condition (Eq. B.1 of Ref. \onlinecite{Amadon2012}).

As the calculation is computationally expensive, we have used different energy windows for the Wannier functions  with increasing width.
We give in Tab. \ref{tab:uo2.static} the variation of the self-consistent $U$ as a function of the 
energy window of the Wannier functions used in the cRPA calculation.  For a large energy window, the value of $U$ and $J$ converge to $U$=5.7 eV and $J$=0.4 eV.

Our values of $U$ and $J$ are thus slightly larger than the commonly used value $U$=4.5 eV and $J$=0.51 eV \cite{Yamazaki1991,Kotani1992}.
Yin {\it et al} \cite{Yin2011} used a related approach\cite{Kutepov2010} to compute the screened interaction
in UO$_2$. They find a value of 6 eV, only slightly larger than ours. 
However, there are no details about the correlated Wannier orbitals used in their calculations. 
The self-consistent $GW$ approach used to compute the interaction might lead to a different band structure
and thus a different screening that in our scheme.
It seems however more consistent to compute the cRPA screened interaction with a DFT+$U$ scheme using the cRPA interaction than a GW scheme using a fully screened interaction.
The rather good agreement with our calculation might possibly comes from the fact that the actual values of the cRPA interaction and the fully screened interaction computed in DFT+$U$ are very close because the $f$ screening is negligible.

To conclude, in order to use the $f$-ext (b) model, which is the most general one,
it is mandatory to carry out the calculation of the cRPA screened interaction starting from a 
DFT+$U$ band structure \cite{Shih2012a,Karlsson2010a} for Mott insulators. It is especially important because most of DFT+DMFT implementations use Wannier orbitals and thus are coherent with a screened interaction computed in this model.

\section{Cerium: Results and discussion}

In this section, we present the static and dynamical cRPA screened interaction in $\alpha$ and $\gamma$ cerium.
The third subsection is devoted to a self-consistent calculation  of the static screened interaction.

\subsection{Static screening}
 \label{subsec:cestatic}
\subsubsection{Bare interaction}
Table \ref{tab:cerium.static} gives the bare interaction as computed in the $f$-ext (b$_1$) model (see Tab. \ref{tab:modelce}).
The values found for $\alpha$ and $\gamma$ cerium are large and in agreement to the values found by Sakuma {\it et al}\cite{Sakuma2013}.
We emphasize that the Wannier functions used in this work are based on the same number of {\it bands} for $\alpha$ and $\gamma$ cerium.
As a consequence, a slightly larger energy window is in fact used for $\alpha$ cerium because the dispersion is more important for a compressed
volume.
If we use the same energy window --- in the $f$-ext (b'$_1$) model, we find instead a value of $V$ for the $\gamma$ phase similar to the value 
found for the $\alpha$ phase. It thus shows that the difference comes from the difference in Wannier functions. 
Anyway, this is a weak effect, at most 2\%.

For $\gamma$ cerium, we compute also the bare interaction in the $f$-ext (b$_2$) and $f$-ext (b$_3$) models with an even more extended window of energy
to construct Wannier functions. We notice an increase of the bare interaction when the energy window is larger, because Wannier functions are more localized.
\label{subsubsec:CeWstatic}
\begin{table}[b]
\centering
\begin{tabular}{ccccccc}
\hline \hline 
	       &   model       &      Bands for        & \multicolumn{2}{c}{$\gamma$}&  \multicolumn{2}{c}{$\alpha$}      \\
	       &               &     Wanniers               &  $U$[eV]    & $J$[eV]   &  $U$[eV]    & $J$[eV]         \\
\hline                                                                              
$v$              & $f$-ext (b$_1$)            &  1-20      &  23.8   & 0.7    &  24.3   &  0.7     \\
$v$              & $f$-ext (b$_2$)            &  1-30      &  25.0   & 0.7    &         &          \\
$v$              & $f$-ext (b$_3$)            &  1-40      &  25.5   & 0.7    &         &          \\
$v$              & $f$-ext (b'$_1$)           &  1-22/20   &  24.2   & 0.7    &  24.3   &  0.7     \\
\hline                                                                              
$W$              & $f$-ext (b$_1$)            &  1-20      &  0.4    & 0.5    &   0.6   &  0.5     \\
\hline                                                                          
$U^{{\rm cRPA}}$ & $f$-ext (b$_1$)            &   1-20     &  0.7    & 0.5    &   0.9   &  0.5     \\
$U^{{\rm cRPA}}$ &  $f$-W$_1$                 &            &  0.5    & 0.5    &   0.7   &  0.5     \\
$U^{{\rm cRPA}}$ &  $fd t_{\rm 2g}$-ext       &  1-20      &  3.8    & 0.6    &   3.8   &  0.6     \\
\hline
	 \multicolumn{7}{c}{LDA+$U$ calculations}     \\
\hline
$U^{{\rm cRPA}}_{\rm  nsc}$ & $f$-ext (b$_1$) &  1-20      &  5.9    & 0.6    &    5.5  &  0.6     \\
$U^{{\rm cRPA}}_{\rm  nsc}$ & $f$-ext (b$_2$) &  1-30      &  6.6    & 0.6    &         &          \\
$U^{{\rm cRPA}}_{\rm  nsc}$ & $f$-ext (b$_3$) &  1-40      &  6.7    & 0.6    &         &          \\
$U^{{\rm cRPA}}_{\rm  nsc}$ & $f$-ext (b$_4$) &  1-50      &  6.7    & 0.6    &         &          \\
$U^{{\rm cRPA}}_{\rm  sc}$ & $f$-ext (b$_1$)  &  1-20      &  5.3    & 0.6    &    0.9  &  0.5     \\
$U^{{\rm cRPA}}_{\rm  sc}$ & $f$-ext (b$_2$)  &  1-30      &  6.5    & 0.6    &    5.4  &  0.6     \\
$U^{{\rm cRPA}}_{\rm  sc}$ & $f$-ext (b$_3$)  &  1-40      &  6.7    & 0.6    &    5.5  &  0.6     \\
$U^{{\rm cRPA}}_{\rm  sc}$ & $f$-ext (b$_4$)  &  1-50      &  6.6    & 0.6    &    5.2  &  0.6     \\
\hline \hline
\end{tabular}
\caption{
Bare ($v$), fully screened ($W$) and cRPA ($U^{\rm cRPA}$) Coulomb interactions 
for different models for cerium in LDA and LDA+$U$ methods. 
The definition of Wannier functions and screening models are defined in Tab. \ref{tab:modelsuo2}.
In  [$f$-ext (b$_i$)]$_{i=1...4}$, the same  number of bands are used to built Wannier functions 
in $\alpha$ and $\gamma$ cerium.
In  [$f$-ext (b'$_1$)], two more bands are used for the calculation of $\gamma$ cerium in order that the same energy window is used in both phases.
The last 8 rows of the table gives value of U obtained from a cRPA calculation starting from a band structure obtained with the LDA+$U$ method.
$U_{\rm  nsc}$ are non self-consistently computed values of $U$, starting from an LDA+$U$ calculation with $U$=6 eV and
$J$=0 eV.
$U_{\rm sc}$ are values of $U$ computed self-consistently with a given energy window to define the Wannier functions.
The self-consistent calculation would be completely coherent when the energy window to define Wannier functions is large.
}
\label{tab:cerium.static}
\end{table}

\subsubsection{Screened interaction}

We discuss now the value of $W$ and $U$ for a fixed volume. 

As shown on Tab. \ref{tab:cerium.static}, they are both small and their small difference is only 
due to $f-f$ transitions. These transitions thus contribute to a reduction
of 0.3 eV of $U$ --- for both phases.
In order to understand the origin of this small value of $U$ and $W$, we removed the transitions from all the $f$ bands to the three bands than are just above 
(mainly with $d t_{\rm 2g}$  character: see Fig. \ref{fig:bsV}). This is an approximated removal because bands are entangled. It corresponds to the
$fd t_{\rm 2g}$ model in Tab. \ref{tab:modelce}.
We find that removing all these screening channels increases the value of the screened interaction from 0.7 eV to 3.8 eV.
So $f-d$ transitions are a major source of screening.
We note that if we had removed only the $f$ bands --- which corresponds to the $f$-W$_1$ model  ---, the value of $U$ would have been small and not far from the value obtained in
the $f$-ext (b$_1$) model (see Tab. \ref{tab:cerium.static}). The comparison of $U$ computed in the $f$-W$_1$ model and the $fd t_{\rm 2g}$ model unambigously demonstrate the important role
of the $f-d t_{\rm 2g}$ transitions.

Nevertheless, for large volume, the appearence of $f$ bands at the Fermi level is in disagreement with experimental photoemission
spectra\cite{Weschke1991,Grioni1997}. This has been largely discussed in the literature (e.g Refs. \cite{Held2001,Amadon2006}).
We carried out the calculation with the LDA+$U$ approximation with  $U$=6 eV and $J$=0 eV\cite{Amadon2008a}. It opens a gap inside the $f$ orbitals, and pushes them apart  
from the Fermi level. Consequently, the screening processes associated with the $f$ orbitals lose weight and
thus the value of the screened interaction is much larger. A similar effect was observed by Karlsson {\it et al} \cite{Karlsson2010a} for gadolinium.
Importantly, the calculated value of $U$ depends largely on the energy window, as shown by the results  ($U^{{\rm LDA+}U}_{\rm  nsc}$  in Tab. \ref{tab:cerium.static}) obtained using the {$f$-ext (b$_i$)}$_{i=1,2,3,4}$  models.
 
We also note as underlined by Sakuma {\it et al},  that $5s$ and $5p$ orbitals contribute much to the screening. Without
their inclusion, the cRPA LDA+$U$ value for the screened interaction would be about 2 eV larger.

Nilsson {\it et al}\cite{Nilsson2013} have 
computed the LDA cRPA screened interaction for $\alpha$ and $\gamma$ cerium.
They 
compute the polarizability of the model 
using a disentangled band structure. Their disentanglement relies on removing the coupling between $f$ and other orbitals.
As a consequence, the screening is reduced\cite{Nilsson2013} and their value of $U$, computed from the LDA
band structure, is larger than ours.
The self-consistency over $U$ might however resolve the discrepancy between the two methods.

\subsubsection{Screened interaction variation as a function of volume}

\begin{figure}[b]
\centering
\begin{tabular}{c}
 {\resizebox{8cm}{!}{\includegraphics{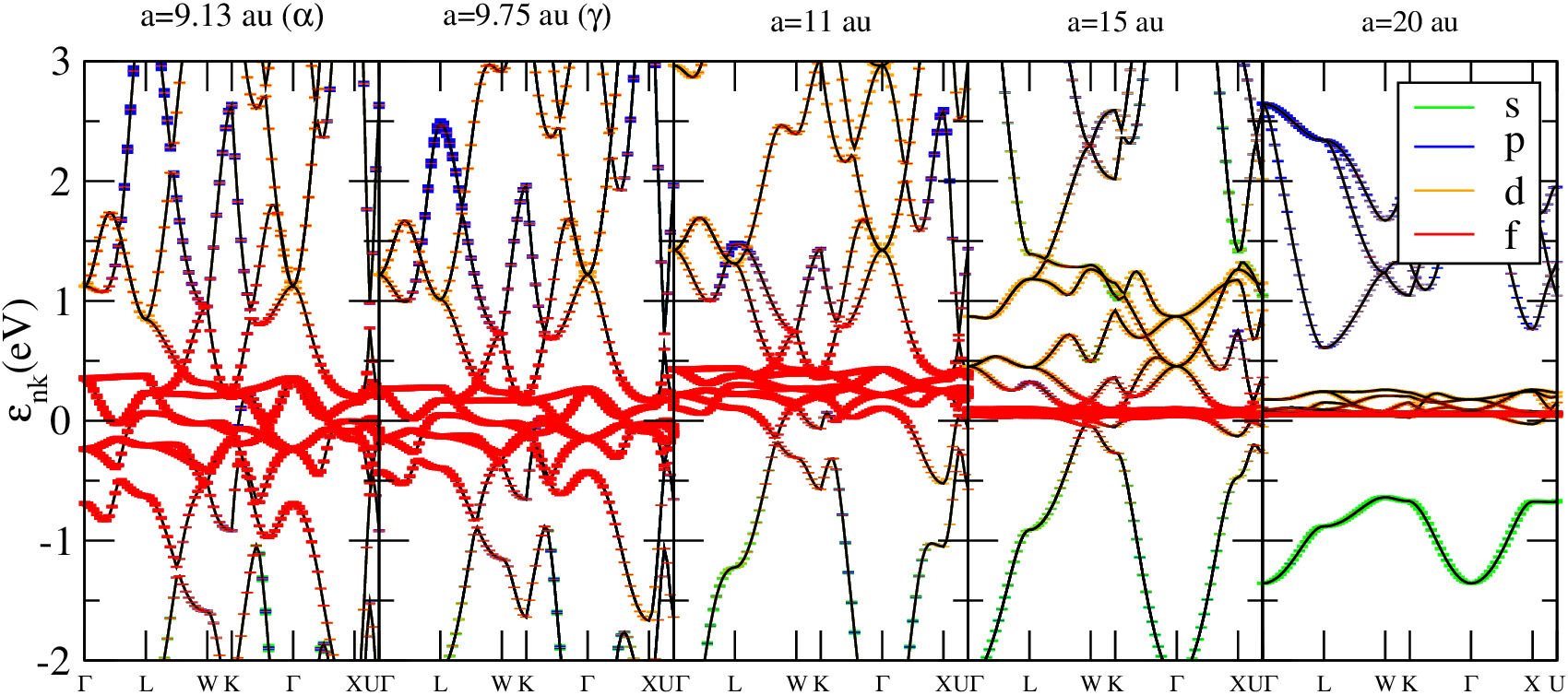}}}
  \\
 {\resizebox{8.5cm}{!}{\includegraphics{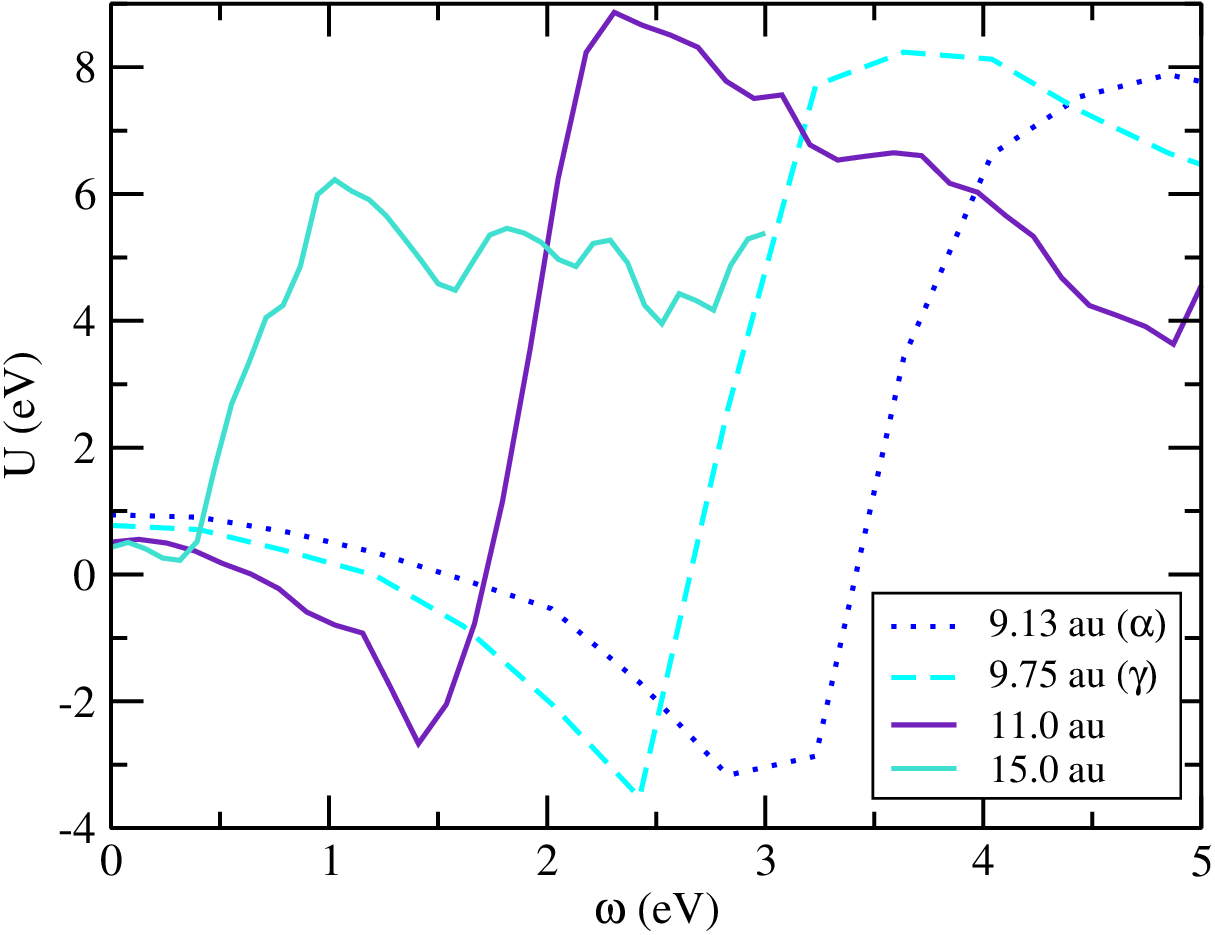}}}
\end{tabular}
  \caption{(Top) Band structure of fcc cerium for different volumes. The characters of $s$, $p$, $d$ and $f$ orbitals
  are indicated by colored fatbands. (Bottom) Screened Coulomb interaction as a function of frequency for different volumes.}
  \label{fig:bsV}
  \label{fig:UWace}
\end{figure}

We now compare the variation of $W$ and $U$ between the two phases.
As discussed above, in subsection \ref{subsubsec:CeWstatic}, Wannier functions in the $f$-ext (b) model
are more localized in the $\alpha$ phase because of the larger dispersion.
So it should induce also a increase of 2\% in $W$ and $U$ .
The differences between the screened interaction in the $\alpha$ phase and in the $\gamma$ phase
is however surprisingly much larger than 2\% and are respectively, of 50 \% and 30\% for $W$ ands $U$.
As seen in Tab. \ref{tab:modelce} $f-d$ transitions are mainly responsible for this. 
Indeed, the calculated value in the $fd t_{\rm 2g}$ model are such that $U_\gamma=3.84$ eV $> U_\alpha=3.79$ eV.
It can be understood from the evolution of the band structure as a function
of the volume of cerium,  as plotted in Fig. \ref{fig:bsV}. When the volume increases, the $f$ and $d$ levels get closer to the Fermi level.
As a consequence, the screening corresponding to the transitions from $f$ to $d$ is more effective.

The same effect holds for the LDA+$U$ approximation. Indeed the cRPA screened interaction is now larger in the $\gamma$ phase as expected: all the low energy screening channels involving $f$ levels are now weaker because they are away from the Fermi level.

\subsection{Dynamical screening}

\begin{figure}[b]
\centering
\begin{tabular}{cc}
 {\resizebox{8cm}{!}{\includegraphics{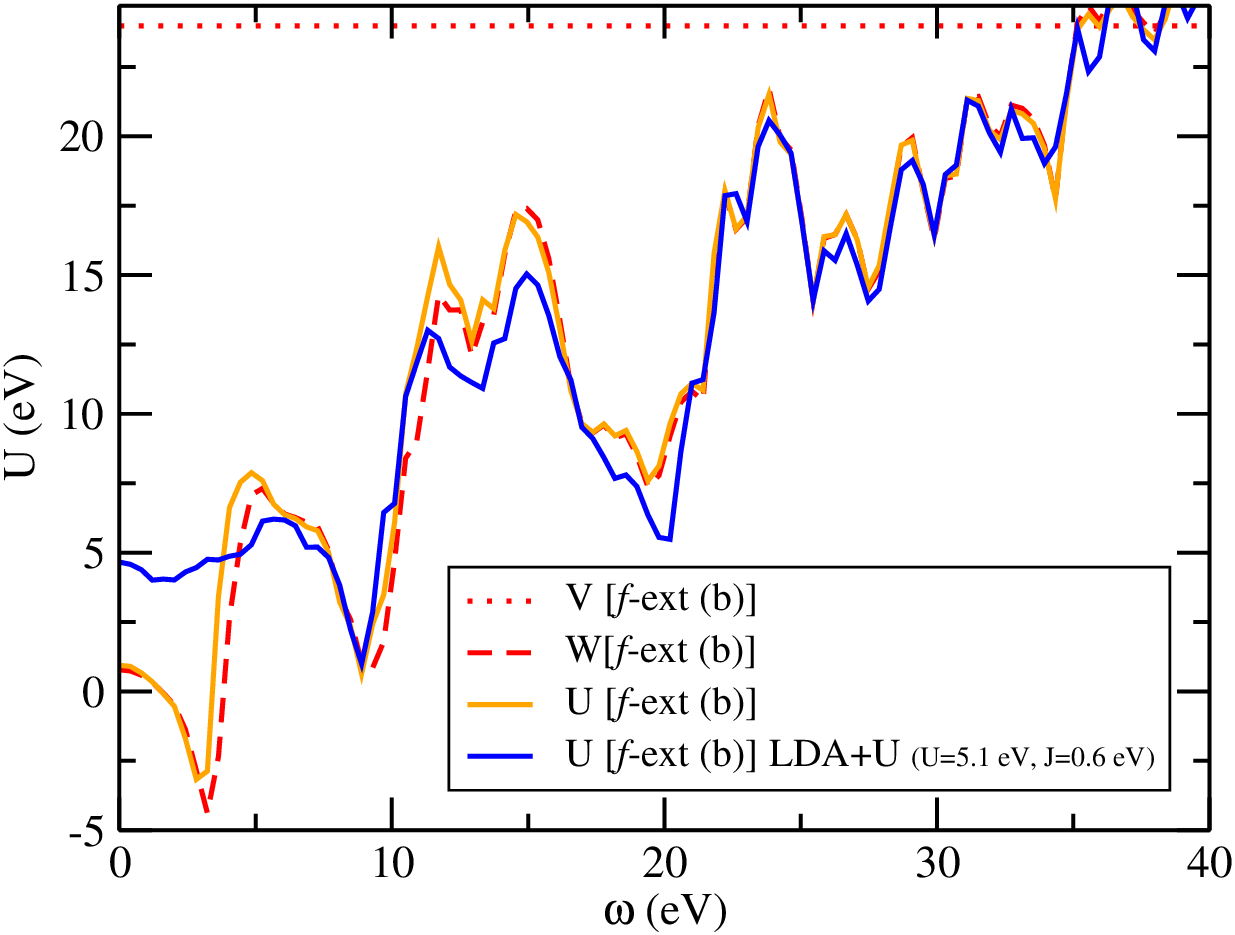}}}
 \\
 {\resizebox{8cm}{!}{\includegraphics{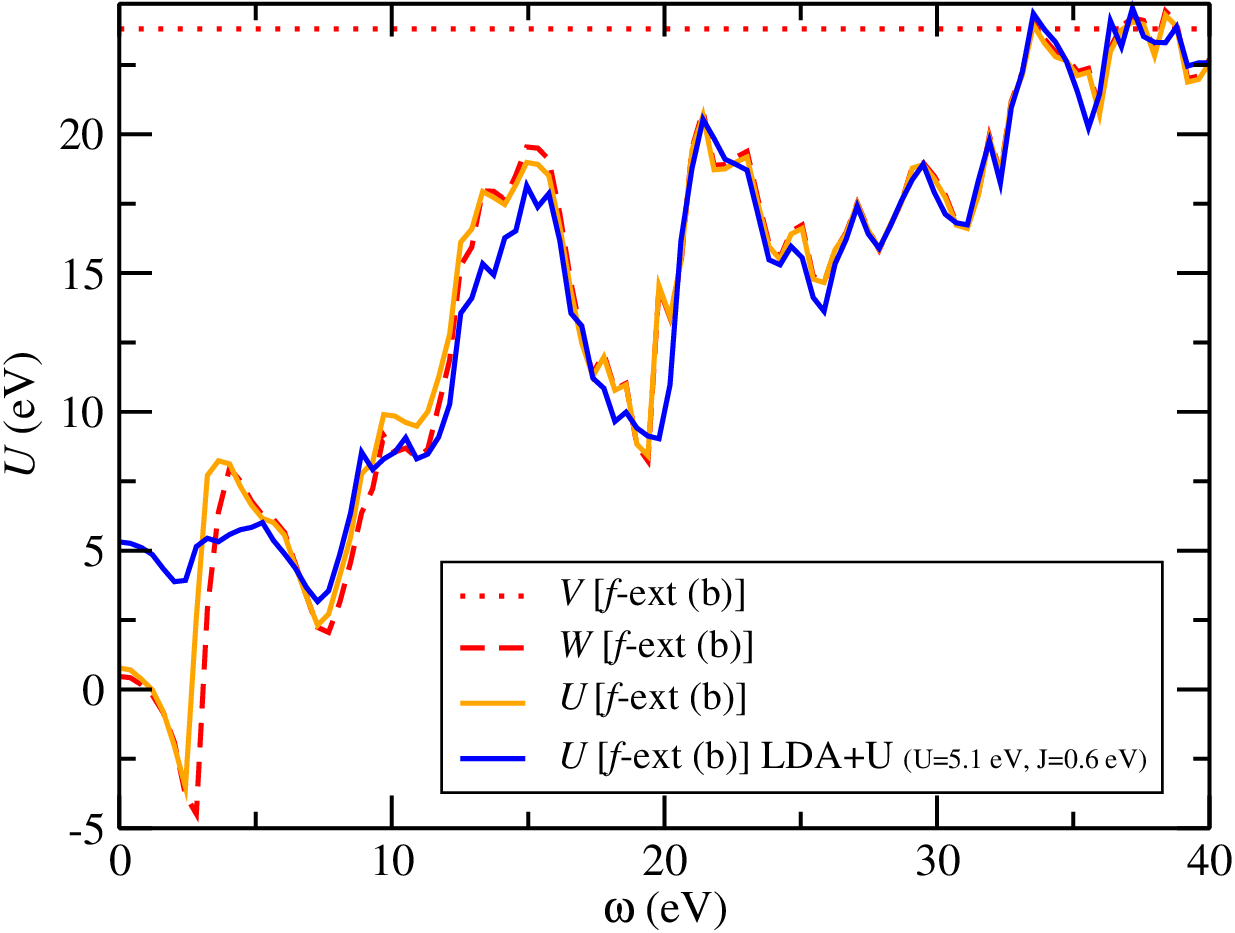}}}
\end{tabular}
\caption{Bare, Fully screened and LDA/LDA+$U$ cRPA partially screened dynamical interactions for $\alpha$ cerium (top) and
$\gamma$ cerium (bottom).}
  \label{fig:ceauw}
 \label{fig:ceguw}
\end{figure}

We plot in Fig. \ref{fig:ceauw} the dynamical screening computed with different models
for $\alpha$ and $\gamma$ cerium.
First of all, the fully screened interaction exhibits three peaks located at 3, 11 and 21 eV --- for $\alpha$ cerium ---, in good agreement to the results of Sakuma {\it et al} \cite{Sakuma2012}. The peak at 21 eV comes from transition from the localized 5$p$ states
as can be checked by removal of these bands from the calculation of the polarizability.
Below 3 eV, both the fully screened and the cRPA interaction --- as computed in LDA --- are weak (below 2 eV). Above 3 eV,
the Coulomb interaction becomes larger.
It can be understood by looking at results from LDA+$U$ calculation. In this calculation, one $f$ band is pushed 2 eV below the Fermi level whereas
the other bands are pushed 2 eV above the Fermi level\cite{Amadon2008a}. The corresponding cRPA screened interaction does not show anymore a peak at
3 eV, because it was originating from transitions involving $f$ orbitals near the Fermi level.

We plot on Fig. \ref{fig:UWace} the evolution of the cRPA interaction as a function of frequency for different volumes.
All curves have the same overall shape: a first domain (a) where the screened interaction is weak (around 2 eV) then a domain (b) where the interaction is larger (around 6 eV).
The width of the domain (a) decreases with the increase of the volume. This is coherent with the argumentation outlined above in subsection \ref{subsec:cestatic}.
The screening which creates the domain (a) 
is due to the proximity of $f$ and $d$ (mainly $t_{\rm 2g}$) states  near the Fermi level. So this screening channel is effective
only for a frequency lower than the $fd$ bandwidth.
As shown in Fig. \ref{fig:bsV}, as the volume increases, this bandwidth decreases in agreement with 
the evolution of the size of domain (a)

\subsection{Self-consistent calculation of $U$ }

As for UO$_2$, it is physically sounded to carry out a self-consistent calculation of $U$  especially in the $\gamma$ phase
because the LDA+$U$ spectral function is qualitatively in good agreement with the photoemission spectra\cite{Amadon2008a}.
So we can expect a better description of screening.

As discussed above, the coherence of the basis for the LDA+$U$ calculation and the Wannier functions has to be preserved.
We thus choose to compute the screened interaction in an atomic basis\cite{Amadon2008a}. As a consequence, a large window
for the Wannier function have to be used\cite{Amadon2012}. We thus carried out the calculation for windows with increasing widths as
shown on Tab. \ref{tab:cerium.static}. 
For the $\gamma$ and $\alpha$ phases, $U$ converges at a value of 6.6 eV and 5.2 eV.
The value can be trusted for the $\gamma$ phase, because this phase is rather well described by LDA+$U$.

However for the $\alpha$ phase, such description is no more valid, because the photoemission spectra\cite{Weschke1991,Grioni1997} exhibits both
a large quasiparticle peak at the Fermi level and Hubbard bands (see e.g Refs. \onlinecite{Bieder2013a}). As a consequence, neither LDA nor LDA+$U$ are able
to describe the correct electronic structure. It has direct implication for the calculation of $U$:
A more correct description of $U$ in $\alpha$ cerium should be carried out with a method which correctly computes the spectral function,
such as DFT+DMFT. We leave it for a future study.

\section{Conclusion}

We report an implementation of the cRPA method in the PAW based DFT/$GW$ code \textsc{Abinit}\cite{Torrent2008,Gonze2009,Giantomassi2009,Giantomassi2011} using Wannier orbitals.
We show the application of the cRPA method to uranium oxide, a Mott insulator, and $\alpha$ and $\gamma$ cerium.
We find that the dynamical screened interactions are particularly peaked because of interband
transitions. Our main results is that an accurate calculation of $U$ for UO$_2$ and $\gamma$ cerium 
can only be obtained by a self-consistent procedure with a coherent choice
of Wannier orbitals.  We show results of the self-consistent calculation to a static $U$ using the DFT+$U$ method. 
For $\alpha$ cerium, we underline that a dynamical calculation of $U$  would be necessary and 
could be obtained by including the screening as described in DFT+DMFT.

\appendix
\section{Expression of $U$}
\label{sec:Udetails}
This appendix gives the expression of $U$ as a function
of oscillator strengh $M_{\vG}^{n, n'}(\vq, \vk)$.
From Eq. \ref{eq:ummmm}, one has the following expressions for $U$\cite{Vaugier2012,Shih2012,Nomura2012} and
the oscillator matrices $M$:
\begin{eqnarray}
\nonumber
U_{m_1, m_2, m_3, m_4}(\omega)
&=&
\frac{1}{\Omega_{cell}}
\sum_{\vq}
w(\vq)
\sum_{\vG\vG'}
M_{\vG}^{m_3, m_1 \dag}(\vq) \\
&&
\times M_{\vG'}^{m_2, m_4}(\vq)
\frac{4\pi\epsilon_{\vG\vG'}^{-1}(\vq,\omega)}{|\vq+\vG'|^2} \\
\nonumber
M_{\vG}^{m, m'}(\vq)&=&
\frac{1}{N_k}\sum_{\vk,n,n'}
M_{\vG}^{n, n'}(\vq, \vk)
C_{m}^{n \vk-\vq \dag}
C_{m'}^{n'\vk} \\
M_{\vG}^{n, n'}(\vq, \vk) &=&
\langle\Psi_{\mathbf{k-q}n}| e^{-i\mathbf{(q+G)r}}| \Psi_{\mathbf{k}n'}\rangle 
\label{eqTW}
\end{eqnarray}

$C_{m}^{n\vk}$ is the coefficient of the expansion of a Wannier function on a Kohn Sham orbital,
$w(\vq)$ is the weight of the $\vq$ vector to sample the Brillouin Zone, $N_k$ is the number of $\vk$ vectors 
in the Brillouin Zone and $\Omega_{cell}$ is the cell volume.
The calculation of the dielectric matrix is detailed in Refs. \onlinecite{Giantomassi2009} and \onlinecite{Giantomassi2011}.
An important point is that in the calculation of both the dielectric matrix and the screened interaction,
the calculation of the oscillator matrix elements $M_{\vG}^{n, n'}(\vq, \vk)$ in the PAW formalism is required.
For these two calculations, we use the scheme of Arnaud and Alouani\cite{Arnaud2000}, as implemented \cite{Giantomassi2009,Giantomassi2011} in \verb?ABINIT?\cite{Gonze2009}. It results from a direct application of Eq. (11) of Ref. \onlinecite{Blochl1994}
to Eq. (\ref{eqTW}).
As underlined in Ref. \onlinecite{Shishkin2007},
it might require a high cutoff energy to compute the dielectric matrix, because it involves the Fourier transform of a product
of the atomic wavefunctions, which are particularly localized for cerium.
We converged the projector basis in order to obtain accurate results.
This was checked in particular by the weak dependence on the sphere radius of the PAW atomic data (at most 0.1 eV).
\section{Benchmark of the implementation on SrVO$_3$}

\label{sec:appendix} 

This appendix gives a comparison of bare and screened interaction for SrVO$_3$  between our calculation
and results obtained with the FLAPW basis by Vaugier {\it et al}\cite{Vaugier2012} and with the PAW basis
by Nomura {\it al} \cite{Nomura2012}.

For the PAW calculations, the atomic data detailed in Ref. \onlinecite{Amadon2008} are used.

The energy cutoffs for the wavefunction, the dielectric function and the calculation of the bare interaction are
15, 7 and 35 Ha. We use a 6x6x6 $k$-mesh grid. All these parameters are sufficient to have a precision better than 0.1 eV on $U$ and $J$.

For all cRPA interactions, we find 
a difference of at most 0.2 eV (6\%)
between our calculation and results from Vaugier {\it et al} and Nomura {\it et al}.
This good agreement gives a further validation of our implementation.

\begin{table}[b]
\centering
\begin{tabular}{ccccc}
\hline \hline 
	      &   model                     &  Ref.\onlinecite{Vaugier2012}& Ref.\onlinecite{Nomura2012}& This work    \\
\hline
$v_{\rm diag}$     & $t_{2g}-t_{2g}$        &   16.1                       & 16.0                       & 16.1         \\
$U_{\rm diag}$     & $t_{2g}-t_{2g}$        &   3.2                        & 3.4                        & 3.4         \\
\hline
$v$                & $d-dp$ (a)             &   19.5                       &                            & 19.4         \\
$U$                & $d-dp$ (a)             &   3.2                        &                            & 3.3         \\
$U_{\rm diag}$     & $d-dp$ (a)             &   4.1                        &                            & 4.3         \\
\hline
$U$                & $dp$                   &   9.9                        &                            & 10.1         \\
\hline \hline
\end{tabular}
\caption{
Bare ($v$), and cRPA ($U$) Coulomb interactions for SrVO$_3$ computed for different models compared to similar calculations
of Ref. \onlinecite{Vaugier2012} and Ref. \onlinecite{Nomura2012}. Notations for the models are taken from Ref. \onlinecite{Vaugier2012} and \onlinecite{Sakuma2013}.}
\label{tab:srvo3.static}
\end{table}

\section{Role of the double counting correction in the cRPA calculation using a DFT+$U$ bandstructure}
\label{app:dc} 

Whereas all results in Tab. \ref{tab:uo2.static} and \ref{tab:cerium.static} are obtained 
with the FLL double counting correction\cite{Anisimov1991,Liechtenstein1995,Czyzyk1994}, we give in this 
appendix results obtained with the Around Mean Field (AMF) double counting correction\cite{Czyzyk1994}. 
This is {\it a priori} less justified than the FLL double counting correction because in our DFT+$U$ calculations
electrons are not delocalized among all f-orbitals.
For uranium dioxide (resp. $\gamma$ cerium) using the $f$-ext (b$_a$) (resp. $f$-ext (b$_1$)) model , we find 
$U^{\rm cRPA}_{\rm  nsc}$ = 4.5 eV (resp. 4.2 eV) instead of 5.0 eV  (resp. 5.9 eV) for the FLL double counting. 

These differences can be understood from the analytical expression of the DFT+$U$ Kohn Sham potential in AMF
(see e.g. Ref. \onlinecite{Amadon2008a}): the Hubbard bands -- and especially the minority spin bands -- are lowered in energy with respect to FLL. It can thus be expected that the contribution of upper Hubbard bands 
to the screening processes is larger, hence the lower value of $U$.

For cerium (with only one correlated electron), the shift 
of Hubbard bands is -1.5 eV, and even larger for the minority bands.
The value of $U$ is thus greatly reduced.
Moreover, the cerium DOS hence obtained in AMF would badly compare with 
experimental photoemission spectra\cite{Grioni1997,Weschke1991}.
As the FLL DOS is better, we can roughly expect that the value of $U$ obtained in FLL is better.
However, and more generally, this comparison calls for a more general self-consistent scheme with a 
more justified double counting correction such as in the GW+DMFT\cite{biermann2003,Sun2004} scheme.

\acknowledgments
We thank Gabriel Antonius, Jordan Bieder, Silke Biermann, Matteo Giantomassi, Fran{\c c}ois Jollet, Cyril Martins, 
Priyanka Seth and Marc Torrent for discussions about this work.
We acknowledge PRACE for awarding us access to  
resource Marenostrum III based in Spain at BSC.
This work was granted access to the HPC resources of CCRT and CINES under the
allocation 2012096681 made by GENCI
(Grand Equipement National de Calcul Intensif).\\

\end{document}